\documentclass[%
reprint,
superscriptaddress,
a4paper,
longbibliography,
amsmath,amssymb,
aps,prl,
numbers,
]{revtex4-1}

\usepackage[pdftex]{hyperref,graphicx}
\usepackage[usenames, svgnames]{xcolor}
\usepackage{amsfonts,amssymb,amsmath,mathrsfs}
\usepackage[separate-uncertainty=true]{siunitx}
\usepackage[english]{babel}
\usepackage[utf8]{inputenc}
\usepackage[T1]{fontenc}
\usepackage{xspace}
\usepackage[normalem]{ulem}

\newcommand{\figref}[2]{\hyperref[#1]{\getrefnumber{#1}(#2)}}

\hypersetup{
	pdftitle = {Interferometric near-field characterization of plasmonic slot
		waveguides in single- and poly-crystalline gold films},
	pdfauthor = {Stefan Linden},
	colorlinks=true,linkcolor=blue,citecolor=blue,filecolor=blue,urlcolor=blue,pdfpagemode=UseNone,pdfstartview={XYZ null null 1.00}
}

\addto\captionsenglish{}
\addto\captionsenglish{}

\renewcommand\textemdash{\leavevmode\unskip\kern0.8pt\rule[0.19\baselineskip]{8pt}{0.4pt}\kern1pt\ignorespaces}

\usepackage[normalem]{ulem}
\usepackage{upgreek}

\begin{document}
\title{Interferometric near-field characterization of plasmonic slot
	waveguides in single- and poly-crystalline gold films}

\author{M. Pr\"amassing}	
\affiliation{Physikalisches Institut, Nussallee 12, Universit\"at Bonn, 53115 Bonn, Germany}

\author{M. Liebtrau}	
\affiliation{Physikalisches Institut, Nussallee 12, Universit\"at Bonn, 53115 Bonn, Germany}	
	
\author{ H.J. Schill}
\affiliation{Physikalisches Institut, Nussallee 12, Universit\"at Bonn, 53115 Bonn, Germany}

	\author{ S. Irsen}
	\affiliation{Center of advanced european studies and research (caesar), Ludwig-Erhard-Allee 2,
		53175 Bonn, Germany.}
	\author{S. Linden}
	\email{linden@physik.uni-bonn.de}
\affiliation{Physikalisches Institut, Nussallee 12, Universit\"at Bonn, 53115 Bonn, Germany}

	\date{\today}






\begin{abstract}
Plasmonic waveguides are a promising platform for integrated nanophotonic circuits and nanoscale quantum optics. Their use is however often hampered by the limited propagation length of the guided surface plasmon modes. A detailed understanding of the influence of the material quality and the waveguide geometry on the complex mode index is therefore crucial. In this letter, we present interferometric near-field measurements at telecommunication wavelength on plasmonic slot waveguides fabricated by focused ion beam milling in  single- and  poly-crystalline gold films. We observe a significantly better performance of the slot waveguides in the single-crystalline gold film for slot widths below $100\,\mathrm{nm}$. In contrast for larger slot widths, both gold films give rise to comparable mode propagation lengths. Our experimental observations indicate that the nature of the dominant loss channel changes with increasing gap size from Ohmic to leakage radiation. Our experimental findings are reproduced by three dimensional numerical calculations.
\end{abstract}
\maketitle

Plasmonic waveguides can support guided surface plasmon polaritons (SPPs) with deep-subwavelength mode cross-sections \cite{gramotnev2010plasmonics}. This feature distinguishes plasmonic waveguides for applications in which the tight lateral confinement of the electromagnetic energy is paramount. For instance, the possibility to undercut the diffraction limit imposed on dielectric waveguides makes their plasmonic counterparts attractive for integrated nanophotonic circuits. In this context, a number of components based on plasmonic waveguides have been recently demonstrated, e.g., nanolasers\cite{oulton2009plasmon}, modulators\cite{melikyan2014high}, directional couplers\cite{gramotnev_directional_2008,thomaschewski2019chip} and nanodetectors\cite{thomaschewski2019chip}.
Plasmonic waveguides are also a promising platform for nanoscale quantum optics. Here, one exploits the enhancement of the local electromagnetic density of states provided by plasmonic waveguides\cite{jun_nonresonant_2008,chang2007strong}.
An excited quantum emitter deposited in the close vicinity of a suitable plasmonic waveguide can emit with high efficiency into the guided SPP mode\cite{akimov2007generation}. 
A handicap of many plasmonic waveguides that goes hand in hand with the tight transversal field confinement is the limited mode propagation length. Typical experimental values range between several microns and tens of microns for optical frequencies.  
A significant contribution to the damping of the guided SPP modes can be attributed to Ohmic losses inside of the metal.
Other possible loss channels are connected with optical scattering from fabrication imperfections and leakage radiation into the substrate.
A possible solution to overcome the limitations resulting from the rather short propagation lengths is to interface plasmonic waveguides with low loss dielectric waveguides \cite{chen2015efficient,chang2007strong}.

\begin{figure}[t]
	\centering
	\fbox{\includegraphics[width=3.33in]{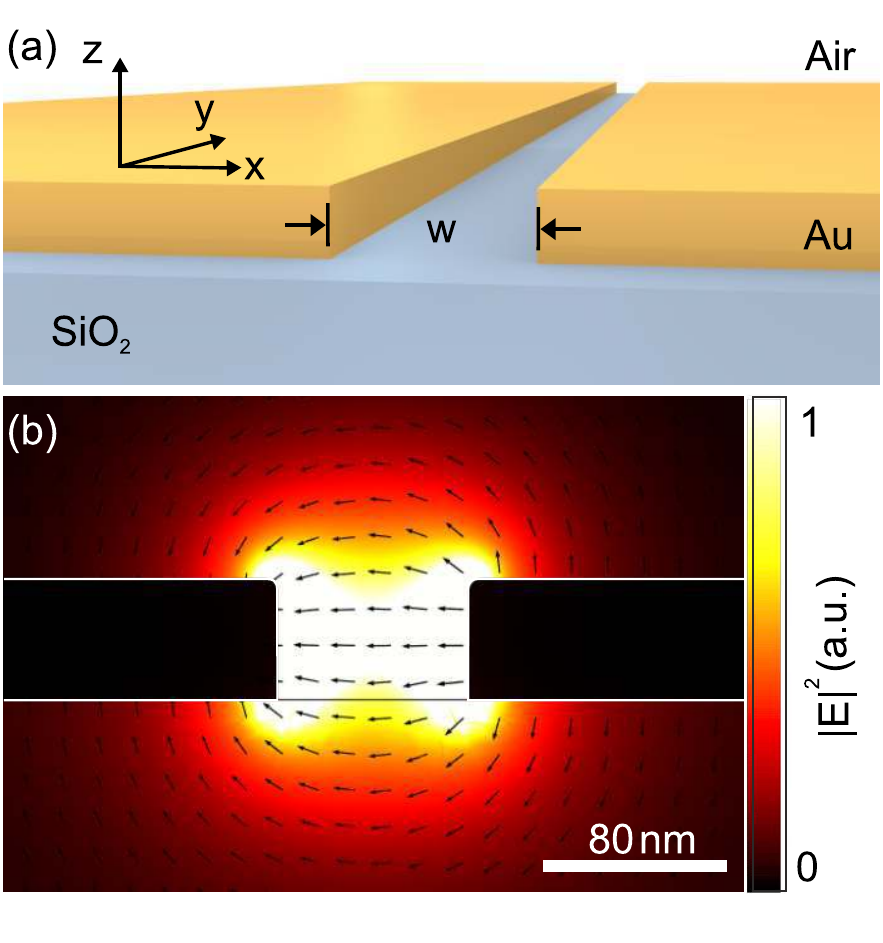}}
	\caption{(a) Schematic representation of a plasmonic slot waveguide. (b) Calculated intensity distribution of the guided plasmonic mode. The black arrows indicate the orientation of the transverse electric field component.}
	\label{fig1}
\end{figure}

In recent years, a wide range of different plasmonic waveguide geometries have been proposed and studied experimentally, e.g., metal nano-wires \cite{akimov2007generation,weeber1999plasmon,krenn2002non}, metal stripes \cite{berini2000plasmon,zenin2016near}, arrays of metallic nano-particles \cite{brongersma2000electromagnetic,maier2003local}, wedges \cite{pile_theoretical_2005}, and v-shaped grooves in metal films \cite{novikov_channel_2002,pile_channel_2004}. One of the most promising configurations with respect to the implementation of planar, on-chip-compatible integrated nanophotonic circuits and opto-electrical interconnects is the plasmonic slot waveguide (PSW)\cite{veronis2005guided,veronis2007modes}. It comprises a narrow, rectangular slot in a thin metal film \cite{gramotnev_directional_2008,melikyan2014high,huang_electrically_2014} and its fundamental SPP mode is strongly confined to the slot region (see Fig.\,\ref{fig1} (a) for a schematic representation of a PSW and (b) for the calculated SPP mode profile). 


Scanning near-field optical microscopy (SNOM) has become an indispensable tool to study the near-field of plasmonic waveguides\cite{bozhevolnyi2006channel,spasenovic2009measurements,andryieuski_direct_2014,zenin2016near,chen2015efficient,zenin2015boosting}. 
In particular, scattering-type SNOM (s-SNOM) in combination with pseudoheterodyne detection is a versatile technique for the characterization of plasmonic waveguides as it allows to measure the amplitude and phase of the waveguide mode from which one can infer its effective index $n$ as well as its propagation length $L$.  
For instance, this technique has been applied to investigate the modes supported by plasmonic strip waveguides\cite{zenin2016near} and to study the coupling of free space radiation to the PSW-mode by different dipole antenna couplers\cite{andryieuski_direct_2014}.


In this letter, we report on the interferometric near-field characterization of strongly confined plasmonic modes supported by PSWs with gap-sizes down to $50\,\mathrm{nm}$ utilizing s-SNOM in transmission mode configuration. The comparison of PSWs defined by focused ion beam milling in chemically synthesized  single-crystalline and thermally evaporated poly-crstalline gold films reveals that the material quality has a significant impact on the mode propagation length for small gap sizes. Surprisingly, the material quality has almost no influence on the mode propagation length for large gap sizes. Our experimental findings are in quantitative agreement with numerical tree-dimensional (3D) finite element calculations. 

\begin{figure}[t]
	\centering
	\fbox{\includegraphics[width=3.33in]{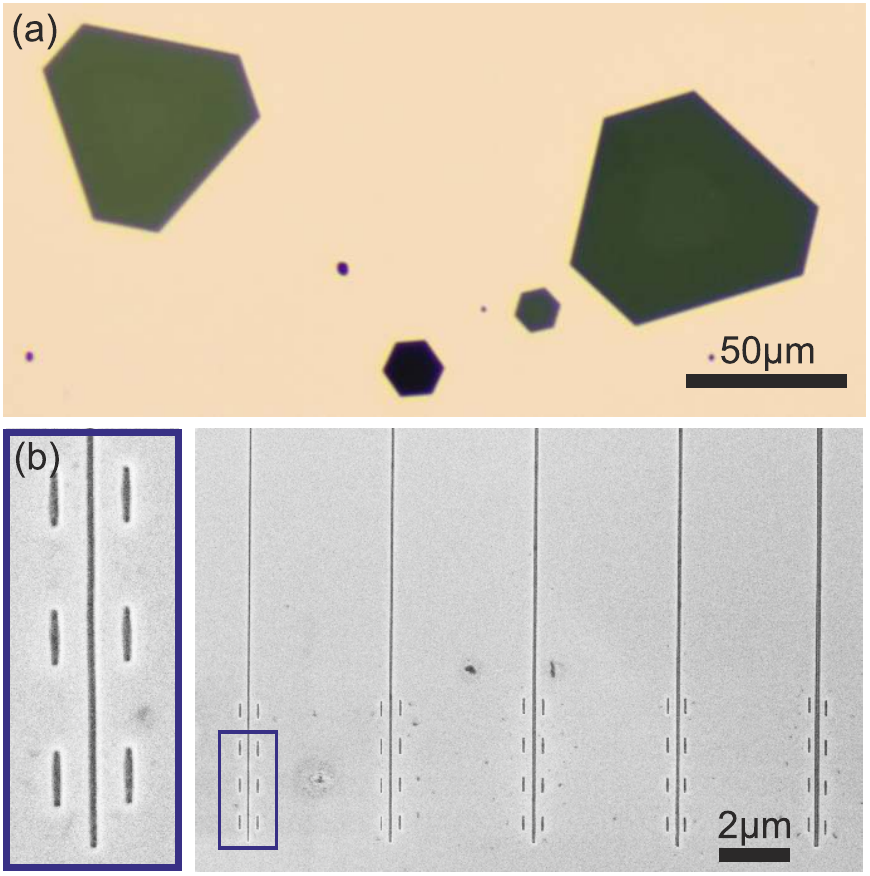}}
	\caption{(a) Optical transmission microscope image of chemically synthesized single-crystalline gold flakes. (b) Scanning electron microscopy images of the investigated PSWs in the single-crystalline film. The blue box displays an enlarged view of the phasematched slot-dipole-antenna array the $50\,$nm PSW.}
	\label{fig2}
\end{figure}

The single-crystalline gold films are fabricated by chemical synthesis in solution\cite{krauss2018controlled}. A mixture of $40\,\upmu$l of a 0.5 molar aqueous solution of chloroauric acid HAuCl$_4$ and $20\,$ml ethyleneglykol is filled into a polypropylene tube. A glass coverslip is immersed into the solution and the tube is heated up to $90\,^\circ$C in an oven for $20\,$h.
Gold atoms released in a three-step chemical reaction can form single-crystalline gold flakes with lateral dimensions larger than $100\,\upmu$m and thicknesses of only a few tens of nanometers (see Fig.\,\ref{fig2} (a)). The glass coverslip is removed from the ethyleneglycol solution and immersed into acetone in order to detach the flakes from the glass coverslip. 
Subsequently, the flakes are transferred by dropcasting from the acetone solution onto a glass substrate, which is coated with a thin layer of Indium Tin Oxide (ITO). 
The electrical conductivity of the ITO layer ensures compatibility of the samples with scanning electron/ion microscopy, while preserving optical transparency for near-infrared wavelengths. We use optical microscopy in combination with atomic force microscopy (AFM) to select a sufficiently large flake with a thickness of $50\,$nm for further processing. For comparison, we also prepare a poly-crystalline gold film by thermal evaporation of a $2\,\mathrm{nm}$-thin chromium adhesion layer and a $50\,\mathrm{nm}$ thick gold film.
In the next step, the PSWs are patterned into the gold films by focused ion beam milling (Zeiss  XB1540 crossbeam). 
We utilize Ga$^{2+}$ ions with an acceleration voltage of $30\,$kV and a beam current of $1.5\,$pA.
Figure \ref{fig2} (b) shows scanning electron micrographs of a set of five waveguides milled in the single crystalline gold film. The slot width $w$ increases from $w=50\,$nm to $w=150\,$nm from left to right. An enlarged view of the end of the $w=50\,$nm PSW is shown in the blue box. 
A phase matched array of four pairs of slot-dipole antennas is milled into the surrounding metal film at the end of each PSW, providing an efficient interface to couple free-space radiation to the waveguide mode. The single slot-antennas are designed to be resonant at telecommunication wavelength and the period of the antenna array matches the slot mode wavelength. 

\begin{figure*}[t]
	\centering
	\fbox{\includegraphics[width=\textwidth]{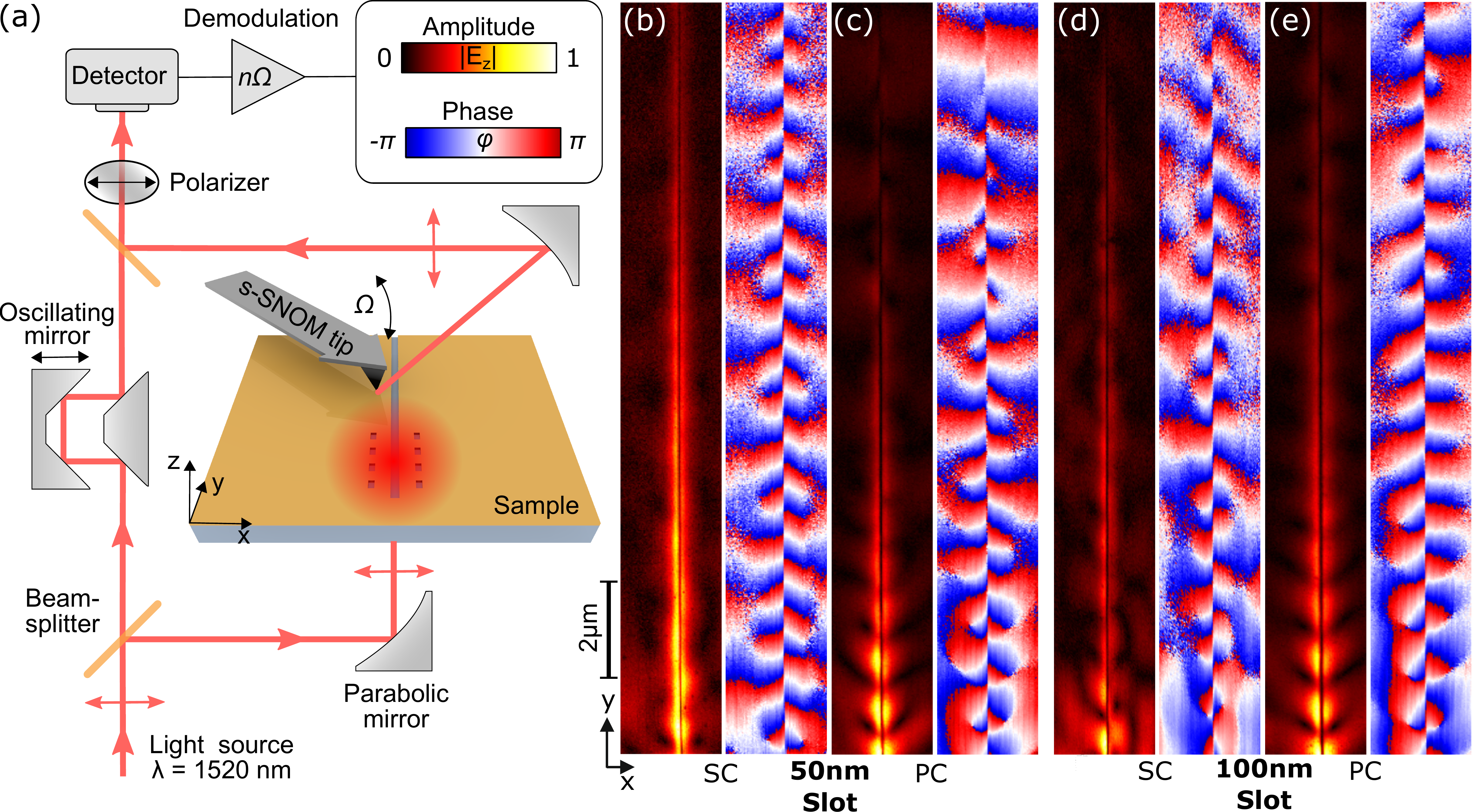}}
	\caption{(a) Scheme of the scattering type scanning near-field optical microscope in transmission mode configuration.. Amplitude $|E_\text{z}|$ and phase $\varphi$ data of the measured near-field distribution of PSWs with different gap widths $w$
		in single-crystalline (SC) and thermally evaporated poly-crystalline (PC) gold films: (b) $w=50\,\mathrm{nm}$, SC; (c) $w=50\,\mathrm{nm}$, PC; (d) $w=100\,\mathrm{nm}$, SC; (e) $w=100\,\mathrm{nm}$, PC. The signals are in each case demodulated at the third harmonic of the tip oscillation frequency $(n=3)$. 
	}
	\label{setup}
\end{figure*}

The near-field of the PSWs is characterized with a s-SNOM (Neaspec neaSNOM) in transmission mode configuration. The setup is schematically depicted in Fig.\,\ref{setup} (a). The attenuated beam of a tunable near-infrared femtosecond light source (Stuttgart Instruments Alpha-HP) tuned to a wavelength of $1520\,\mathrm{nm}$ is spatially filtered by a single mode fiber and focused onto the antenna coupler array from below by a parabolic mirror. The waist radius of the Gaussian beam at the sample surface is determined to be approximately $4\,\upmu$m by a knife edge measurement. The incident beam is polarized along the $x$-direction to resonantly excite the slot-antenna array and efficiently couple to the waveguide mode, which propagates along the slot in $y$-direction. A metallized AFM tip (Nanoworld Arrow NCPt) is approached to the sample by means of tapping mode atomic force microscopy. Due to its elongated shape in vertical direction, the tip acts as an antenna, predominantly scattering the out-of-plane near-field component $E_\text{z}$ at the sample surface to the far-field. The scattered radiation is collimated and guided to an Indium-Gallium-Arsenide detector by the upper parabolic mirror. To measure both the amplitude and phase of the near-field by psydoheterodyne detection\cite{ocelic2006pseudoheterodyne}, the scattered signal is overlapped with a reference beam. Demodulation of the detector signal at a higher harmonics of the tip oscillation frequency $\Omega$ allows to strongly suppress background contributions, e.g. light directly scattered from the sample or the tip.

By raster scanning the sample with respect to the fixed configuration of the tip and the upper parabolic mirror we obtain a two-dimensional (2D) map of the near-field distribution. A synchronized movement of the lower parabolic mirror along with the sample guarantees a fixed illumination of the antenna coupler array during the scan. Movements of the lower parabolic mirror along $x$-direction require a correction of the phase raw-data by an additive contribution to account for the change of the optical path length with respect to the reference arm of the interferometer: $\varphi=\varphi_\text{raw}+2\pi\frac{x}{\lambda}$. Movements along the $y$-direction require no phase correction, since they do not alter the optical path length.

Figure \ref{setup} (b) shows amplitude and phase data of a near-field measurement along a $w=50\,$nm wide PSW in single-crystalline gold, demodulated at the third harmonic of the tip oscillation frequency $3\Omega$. The tapping amplitude of the tip is set to $70\,$nm and the wavelength of the incident laser is $1520\,$nm for all measurements. The antenna array, which excites the waveguide mode is not included in the scan region and lies further below in negative $y$-direction. The near-field amplitude $|E_\text{z}|$ is tightly confined around the slot in $x$-direction and decreases during mode propagation along the $y$-direction. 
As expected from the transverse electric field distribution depicted in Fig.\,\ref{fig1} (b), the measured out-of-plane field component $|E_\text{z}|$ vanishes in the middle of the slot.
The corresponding phase data clearly displays the expected antisymmetric nature of the mode with a phase jump of $\pi$ across the slot. During mode propagating, we observe a linear phase increase in $y$-direction.

The experimental data for the $w=50\,\mathrm{nm}$ wide PSW milled into the poly-crystalline gold film is shown in Fig.\,\ref{setup} (c). It is immediately apparent that the propagation length is here significantly shorter than in the previous case. For that reason, we had to move the scan region closer to the excitation such that the exciting beam partially overlaps with the scan region. Therefore, we have a significant contribution of free propagating SPPs on the gold/air interface excited by the incident beam on the edges of the slot.
The interference of the waveguide mode propagating in $y$-direction and the SPPs running in $\pm x$-direction results in the observed beating pattern\cite{andryieuski_direct_2014}.

Figure \ref{setup} (d) and (e) display the amplitude and phase maps of PSWs with a slot width of $w=100\,\mathrm{nm}$ milled into the single-crystalline gold film and the poly-crystalline gold film, respectively.
In contrast to the case of the $w=50\,\mathrm{nm}$ wide slot, we now observe comparable propagation lengths for both film types. Again, the beating pattern can be attributed to the interference of the slot mode with freely propagating SPPs.


In order to extract the effective mode index $n$ and the propagation length $L$ of the waveguide mode from the measured near-field distributions, we fit the following 2D model, which describes the $E_\text{z}$-component of the waveguide mode outside of the gap region, to the data:
\begin{equation*}
E_\text{z}(x,y)=\begin{cases}
E_r \,e^{[y(\imath k_0 n-\frac{1}{2L})-\frac{x-x_0}{\kappa}+\imath \varphi_0]} & x-x_0 \geq w/2 \\
-E_l \,e^{[y(\imath k_0 n-\frac{1}{2L})-\frac{x_0-x}{\kappa}+\imath\varphi_0] }& x_0-x \geq w/2
\end{cases}
   \label{eqn:fitfun}
\end{equation*}

In this model, the center of the slot is located at $x=x_0$. The gap region is excluded from the fit. 
We allow different values for the amplitudes $E_r$ and $-E_l$ on the right and left hand-side of the gap to account for the slight asymmetry of the near-field amplitude in the measurements, which we attribute to the specific geometry of the detection optics in the s-SNOM. i.e., the relative positions of the slot with respect to the AFM tip and the collecting parabolic mirror.
The opposite algebraic signs take the phase jump of $\pi$ across the slot into account, $\varphi_0$ represents a global phase and $k_0=2\pi/\lambda_0$ is the free space wavenumber. The evanescent decay of the mode away from the slot region is described by the parameter $\kappa$.

\begin{figure}[htbp]
	\centering
	\fbox{\includegraphics[width=3.33in]{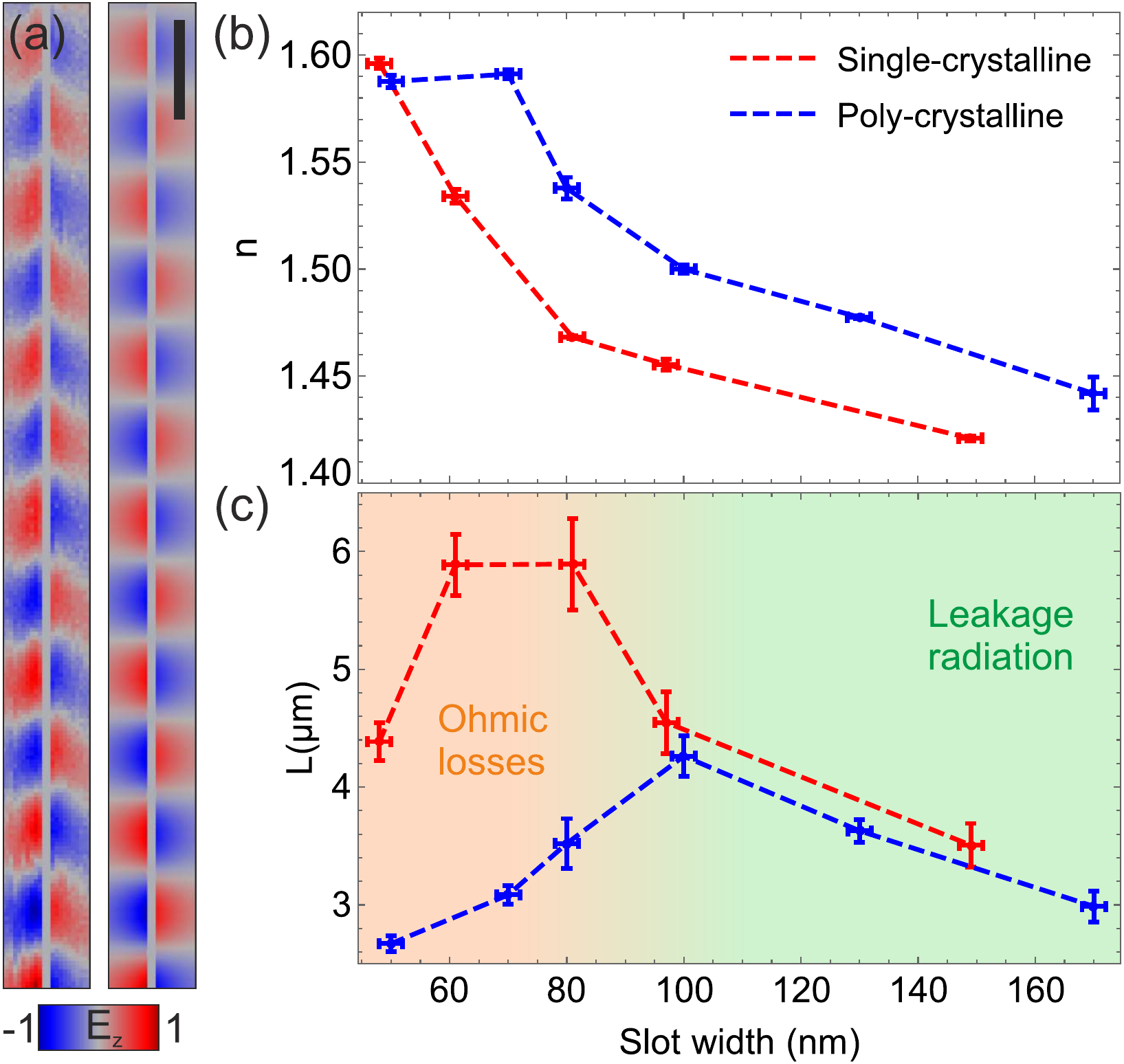}}
	\caption{Results of the 2D complex fits. (a) Real part of the $E_z$ electric field component around the $w=50\,$nm slot in the single-crystalline film: Measured data (left panel) and fit (right panel). The scale bar has a length of $500\,\mathrm{nm}$. Effective mode indices (b) and propagation lengths (c) both on poly-crystalline and single-crystalline gold for different slot width.}
	\label{results}
\end{figure}

Figure \ref{results} (a) exemplarily shows the $E_z$ component of the experimental data and of the fitted 2D model for the case of the $w=50\,$nm wide PSW milled in the single-crystalline film.
The resulting parameters for the propagation length $L$ and the effective mode index $n$ for gap widths are plotted in Fig.\,\ref{results} (b) and (c), respectively. The error bars result from averaging fit results from the measurement data of the third and fourth demodulation order as well as the forward and backward scan direction, respectively. 
For the single- as well as the poly-crystalline gold film, the effective mode index decreases with increasing gap size. This behaviour is expected from theoretical considerations\cite{veronis2007modes} (also see below). In the poly-crystalline film, the mode index is slightly larger than that of the single-crystalline sample. 
We attribute this to the influence of the $15\,$nm thin ITO layer that we had to deposit on the glass substrate in the case of the single-crystalline gold film for fabrication reasons and that is absent in the case of the evaporated gold film.

Interestingly, the propagation length shows for both films a non-monotonic behaviour. In the case of the single-crystalline gold film, we observe a maximum propagation length of $L\approx6\,\upmu\mathrm{m}$ for a gap width of $w=80\,\mathrm{nm}$. The maximum propagation length for the poly-crystalline gold film is about $L\approx4\,\upmu\mathrm{m}$ reached for a gap width of $w=100\,\mathrm{nm}$.
We explain this non-monotonic behaviour by a trade-off between two different loss channels.
With decreasing gap size, a larger fraction of the electric field resides in the metal and the Ohmic losses increase. 
In this region, we expect that the quality of the gold film plays a crucial role.
This is in accordance with our observation that for gap sizes below $w=100\,\mathrm{nm}$ the propagation length of the PSWs defined in the single-crystalline gold film is up to a factor two larger than that of the PSWs in the poly-crystalline gold film.
In contrast for large gap sizes, we expect that the Ohmic losses are reduced since the electric field of the mode is mostly located in the gap region and in the dielectric surrounding of the metal film. Hence, one could naively expect that $L$ should monotonically increase with the gap size.  Nevertheless, we observe a different behaviour, i.e., the propagation length decreases for the largest gap sizes for both gold films. The reason for this is leakage radiation, which becomes the dominant loss channel once the effective mode index deceeds the refractive index of the substrate.
As a result, we observe for both gold films comparable propagation lengths in the case of large gap sizes above $w=100\,\mathrm{nm}$.


\begin{figure}[htbp]
	\centering
	\fbox{\includegraphics[width=3.33in]{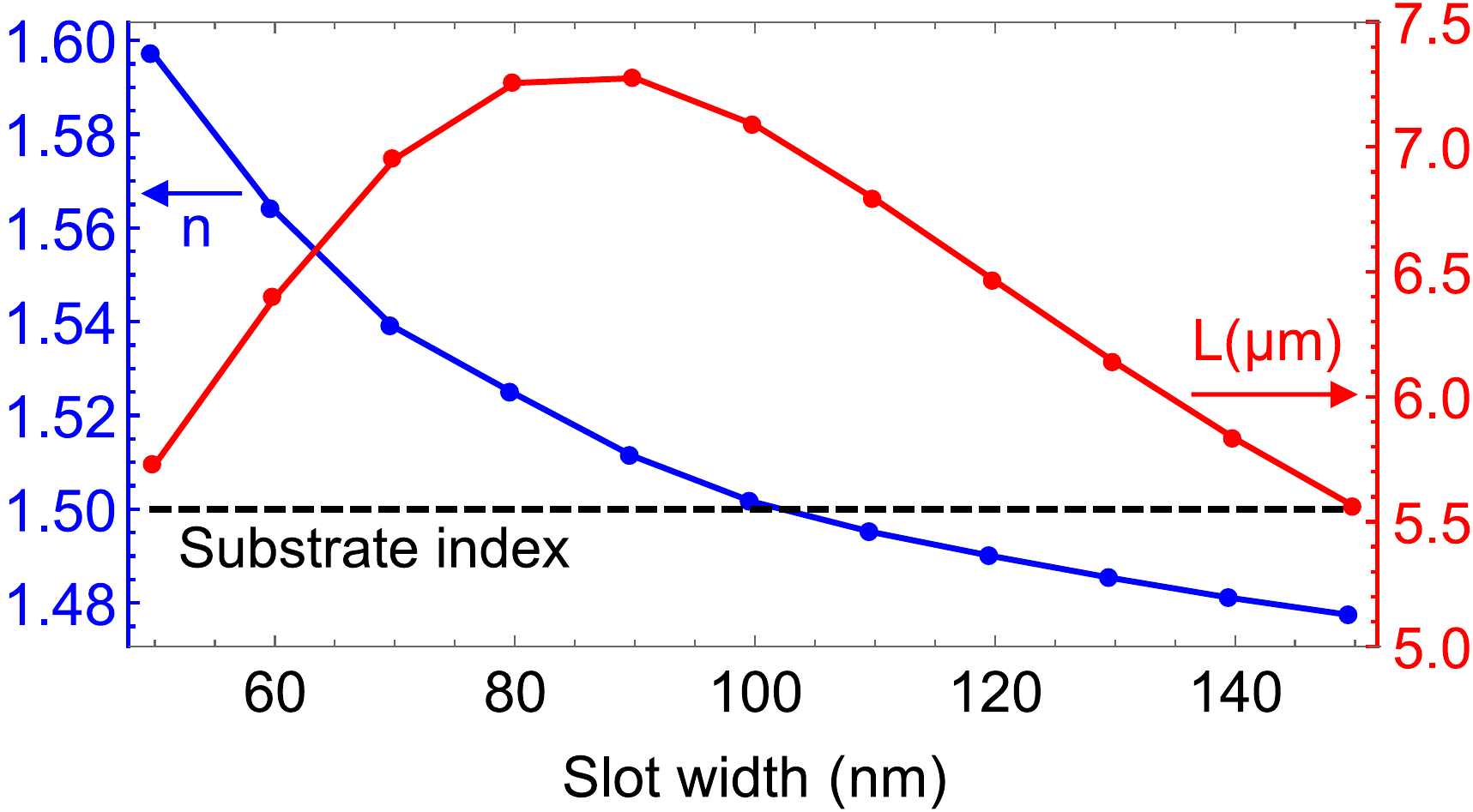}}
	\caption{Calculated effective index (blue curve) and propagation length (red curve) of the PSW slot mode as a function of gap size.}
	\label{Fig_5}
\end{figure}

In order to support this interpretation, we have performed 3D numerical calculations employing a commercial finite element method solver (COMSOL Multiphysics).
The total size of the computational domain is $1.5 \,\upmu\mathrm{m} \times 20\,\upmu\mathrm{m} \times 3 \,\upmu\mathrm{m}$ $(x \times y \times z)$. 
We take advantage of the symmetry of the PSW by applying perfect electric conductor boundary conditions to the $yz$-plane in the middle of the gap. All other boundaries of the computational domain are terminated by perfectly matched layers  with a thickness of $750\,\mathrm{nm}$.  
The $50\,\mathrm{nm}$ gold film is sandwiched between the glass substrate below (refractive index n=1.5) and air (refractive index n=1) above. The gap region consists of air.
For the dielectric function of the gold film, we use the data provided by Johnson and Christy\cite{johnson1972optical}.  
The slot waveguide has a length of $18.5\,\upmu\mathrm{m}$ and is terminated in $y$-direction on one end by $1.5\,\upmu\mathrm{m}$ of gold and on the other end by a perfectly matched layer.
For reasons of calculation time and memory consumption, we omit the phase-matched slot-dipole-antenna array and excite the slot mode at the closed end of the waveguide by a normally incident, $x$-polarized Gaussian beam from the substrate side.
The slot width is varied in a parameter sweep from $w=40\,\mathrm{nm}$ to $w=150\,\mathrm{nm}$ in steps of $10\,\mathrm{nm}$.
From each of these calculations, we extract the transverse electric field component $E_x(y)$ along a line profile through the center of the slot in propagation direction ($y$-direction). The effective mode index and the propagation distance are obtained by fitting a one-dimensional damped harmonic wave to the $E_x(y)$-data (see Fig.\ref{Fig_5}).
The numerical results are in good agreement with the experimental findings for the single-crystalline gold film. In particular, they also show a monotonic decrease of the effective mode index and a maximum of the propagation length for gap sizes around $w\approx 90\,\mathrm{nm}$.
As in the experiment, we observe that the propagation length strongly decreases once the effective mode index becomes smaller than the refractive index of the substrate.
We note that the latter aspect cannot be reproduced with 2D mode analysis calculations, in which leakage of the slot mode into the substrate is not properly taken into account. 

We have fabricated plasmonic slot waveguides in single-crystalline and poly-crystalline gold films with slot widths down to $50\,\mathrm{nm}$. The effective mode index and the propagation length were experimentally determined by amplitude- and phase resolved near-field measurements with a scattering type scanning near-field optical microscope. We find that the quality of the gold film has a profound impact for small gap widths for which Ohmic losses are the dominant loss channel. In particular, we observe that the propagation length can be improved by almost a factor of two by choosing a high-quality single-crystalline gold film. In contrast, for gap widths exceeding $100\,\mathrm{nm}$, the waveguides in single-crystalline and poly-crystalline gold films show a comparable performance since leakage radiation into the substrate becomes the dominant loss channel. However, The latter could be suppressed by embedding the gold film in a symmetric dielectric environment, e.g., by coating the gold film with a spin-on-dielectric\cite{veronis2007modes}.


\begin{acknowledgments}
S.L. and S.I. acknowledge financial support by the Deutsche Forschungsgemeinschaft (LI 1641/5-1).
Furthermore, the authors acknowledge Mario Hentschel from the University of Stuttgart for support regarding the sample fabrication.
\end{acknowledgments}


\bibliography{WGPaperBib}

\begin{thebibliography}{28}%
\makeatletter
\providecommand \@ifxundefined [1]{%
 \@ifx{#1\undefined}
}%
\providecommand \@ifnum [1]{%
 \ifnum #1\expandafter \@firstoftwo
 \else \expandafter \@secondoftwo
 \fi
}%
\providecommand \@ifx [1]{%
 \ifx #1\expandafter \@firstoftwo
 \else \expandafter \@secondoftwo
 \fi
}%
\providecommand \natexlab [1]{#1}%
\providecommand \enquote  [1]{``#1''}%
\providecommand \bibnamefont  [1]{#1}%
\providecommand \bibfnamefont [1]{#1}%
\providecommand \citenamefont [1]{#1}%
\providecommand \href@noop [0]{\@secondoftwo}%
\providecommand \href [0]{\begingroup \@sanitize@url \@href}%
\providecommand \@href[1]{\@@startlink{#1}\@@href}%
\providecommand \@@href[1]{\endgroup#1\@@endlink}%
\providecommand \@sanitize@url [0]{\catcode `\\12\catcode `\$12\catcode
  `\&12\catcode `\#12\catcode `\^12\catcode `\_12\catcode `\%12\relax}%
\providecommand \@@startlink[1]{}%
\providecommand \@@endlink[0]{}%
\providecommand \url  [0]{\begingroup\@sanitize@url \@url }%
\providecommand \@url [1]{\endgroup\@href {#1}{\urlprefix }}%
\providecommand \urlprefix  [0]{URL }%
\providecommand \Eprint [0]{\href }%
\providecommand \doibase [0]{http://dx.doi.org/}%
\providecommand \selectlanguage [0]{\@gobble}%
\providecommand \bibinfo  [0]{\@secondoftwo}%
\providecommand \bibfield  [0]{\@secondoftwo}%
\providecommand \translation [1]{[#1]}%
\providecommand \BibitemOpen [0]{}%
\providecommand \bibitemStop [0]{}%
\providecommand \bibitemNoStop [0]{.\EOS\space}%
\providecommand \EOS [0]{\spacefactor3000\relax}%
\providecommand \BibitemShut  [1]{\csname bibitem#1\endcsname}%
\let\auto@bib@innerbib\@empty
\bibitem [{\citenamefont {Gramotnev}\ and\ \citenamefont
  {Bozhevolnyi}(2010)}]{gramotnev2010plasmonics}%
  \BibitemOpen
  \bibfield  {author} {\bibinfo {author} {\bibfnamefont {Dmitri~K}\
  \bibnamefont {Gramotnev}}\ and\ \bibinfo {author} {\bibfnamefont {Sergey~I}\
  \bibnamefont {Bozhevolnyi}},\ }\bibfield  {title} {\enquote {\bibinfo {title}
  {Plasmonics beyond the diffraction limit},}\ }\href@noop {} {\bibfield
  {journal} {\bibinfo  {journal} {Nature Photonics}\ }\textbf {\bibinfo
  {volume} {4}},\ \bibinfo {pages} {83} (\bibinfo {year} {2010})}\BibitemShut
  {NoStop}%
\bibitem [{\citenamefont {Oulton}\ \emph {et~al.}(2009)\citenamefont {Oulton},
  \citenamefont {Sorger}, \citenamefont {Zentgraf}, \citenamefont {Ma},
  \citenamefont {Gladden}, \citenamefont {Dai}, \citenamefont {Bartal},\ and\
  \citenamefont {Zhang}}]{oulton2009plasmon}%
  \BibitemOpen
  \bibfield  {author} {\bibinfo {author} {\bibfnamefont {Rupert~F}\
  \bibnamefont {Oulton}}, \bibinfo {author} {\bibfnamefont {Volker~J}\
  \bibnamefont {Sorger}}, \bibinfo {author} {\bibfnamefont {Thomas}\
  \bibnamefont {Zentgraf}}, \bibinfo {author} {\bibfnamefont {Ren-Min}\
  \bibnamefont {Ma}}, \bibinfo {author} {\bibfnamefont {Christopher}\
  \bibnamefont {Gladden}}, \bibinfo {author} {\bibfnamefont {Lun}\ \bibnamefont
  {Dai}}, \bibinfo {author} {\bibfnamefont {Guy}\ \bibnamefont {Bartal}}, \
  and\ \bibinfo {author} {\bibfnamefont {Xiang}\ \bibnamefont {Zhang}},\
  }\bibfield  {title} {\enquote {\bibinfo {title} {Plasmon lasers at deep
  subwavelength scale},}\ }\href@noop {} {\bibfield  {journal} {\bibinfo
  {journal} {Nature}\ }\textbf {\bibinfo {volume} {461}},\ \bibinfo {pages}
  {629} (\bibinfo {year} {2009})}\BibitemShut {NoStop}%
\bibitem [{\citenamefont {Melikyan}\ \emph {et~al.}(2014)\citenamefont
  {Melikyan}, \citenamefont {Alloatti}, \citenamefont {Muslija}, \citenamefont
  {Hillerkuss}, \citenamefont {Schindler}, \citenamefont {Li}, \citenamefont
  {Palmer}, \citenamefont {Korn}, \citenamefont {Muehlbrandt}, \citenamefont
  {Van~Thourhout}, \citenamefont {Chen}, \citenamefont {Dinu}, \citenamefont
  {Sommer}, \citenamefont {Koos}, \citenamefont {Freude},\ and\ \citenamefont
  {Leuthold}}]{melikyan2014high}%
  \BibitemOpen
  \bibfield  {author} {\bibinfo {author} {\bibfnamefont {A.}~\bibnamefont
  {Melikyan}}, \bibinfo {author} {\bibfnamefont {L.}~\bibnamefont {Alloatti}},
  \bibinfo {author} {\bibfnamefont {A.}~\bibnamefont {Muslija}}, \bibinfo
  {author} {\bibfnamefont {D.}~\bibnamefont {Hillerkuss}}, \bibinfo {author}
  {\bibfnamefont {P.~C.}\ \bibnamefont {Schindler}}, \bibinfo {author}
  {\bibfnamefont {J.}~\bibnamefont {Li}}, \bibinfo {author} {\bibfnamefont
  {R.}~\bibnamefont {Palmer}}, \bibinfo {author} {\bibfnamefont
  {D.}~\bibnamefont {Korn}}, \bibinfo {author} {\bibfnamefont {S.}~\bibnamefont
  {Muehlbrandt}}, \bibinfo {author} {\bibfnamefont {D.}~\bibnamefont
  {Van~Thourhout}}, \bibinfo {author} {\bibfnamefont {B.}~\bibnamefont {Chen}},
  \bibinfo {author} {\bibfnamefont {R.}~\bibnamefont {Dinu}}, \bibinfo {author}
  {\bibfnamefont {M.}~\bibnamefont {Sommer}}, \bibinfo {author} {\bibfnamefont
  {Kohl~M.}\ \bibnamefont {Koos}, \bibfnamefont {C.}}, \bibinfo {author}
  {\bibfnamefont {W.}~\bibnamefont {Freude}}, \ and\ \bibinfo {author}
  {\bibfnamefont {J.}~\bibnamefont {Leuthold}},\ }\bibfield  {title} {\enquote
  {\bibinfo {title} {High-speed plasmonic phase modulators},}\ }\href@noop {}
  {\bibfield  {journal} {\bibinfo  {journal} {Nature Photonics}\ }\textbf
  {\bibinfo {volume} {8}},\ \bibinfo {pages} {229} (\bibinfo {year}
  {2014})}\BibitemShut {NoStop}%
\bibitem [{\citenamefont {Gramotnev}\ \emph {et~al.}(2008)\citenamefont
  {Gramotnev}, \citenamefont {Vernon},\ and\ \citenamefont
  {Pile}}]{gramotnev_directional_2008}%
  \BibitemOpen
  \bibfield  {author} {\bibinfo {author} {\bibfnamefont {D.~K.}\ \bibnamefont
  {Gramotnev}}, \bibinfo {author} {\bibfnamefont {K.~C.}\ \bibnamefont
  {Vernon}}, \ and\ \bibinfo {author} {\bibfnamefont {D.~F.~P.}\ \bibnamefont
  {Pile}},\ }\bibfield  {title} {\enquote {\bibinfo {title} {Directional
  coupler using gap plasmon waveguides},}\ }\href@noop {} {\bibfield  {journal}
  {\bibinfo  {journal} {Appl. Phys. B}\ }\textbf {\bibinfo {volume} {93}},\
  \bibinfo {pages} {99--106} (\bibinfo {year} {2008})}\BibitemShut {NoStop}%
\bibitem [{\citenamefont {Thomaschewski}\ \emph {et~al.}(2019)\citenamefont
  {Thomaschewski}, \citenamefont {Yang}, \citenamefont {Wolff}, \citenamefont
  {Roberts},\ and\ \citenamefont {Bozhevolnyi}}]{thomaschewski2019chip}%
  \BibitemOpen
  \bibfield  {author} {\bibinfo {author} {\bibfnamefont {Martin}\ \bibnamefont
  {Thomaschewski}}, \bibinfo {author} {\bibfnamefont {Yuanqing}\ \bibnamefont
  {Yang}}, \bibinfo {author} {\bibfnamefont {Christian}\ \bibnamefont {Wolff}},
  \bibinfo {author} {\bibfnamefont {Alexander~Sylvester}\ \bibnamefont
  {Roberts}}, \ and\ \bibinfo {author} {\bibfnamefont {Sergey~I}\ \bibnamefont
  {Bozhevolnyi}},\ }\bibfield  {title} {\enquote {\bibinfo {title} {On-chip
  detection of optical spin-orbit interactions in plasmonic nanocircuits},}\
  }\href@noop {} {\bibfield  {journal} {\bibinfo  {journal} {Nano Lett.}\ }
  (\bibinfo {year} {2019})}\BibitemShut {NoStop}%
\bibitem [{\citenamefont {Jun}\ \emph {et~al.}(2008)\citenamefont {Jun},
  \citenamefont {Kekatpure}, \citenamefont {White},\ and\ \citenamefont
  {Brongersma}}]{jun_nonresonant_2008}%
  \BibitemOpen
  \bibfield  {author} {\bibinfo {author} {\bibfnamefont {Y.~C.}\ \bibnamefont
  {Jun}}, \bibinfo {author} {\bibfnamefont {R.~D.}\ \bibnamefont {Kekatpure}},
  \bibinfo {author} {\bibfnamefont {J.~S.}\ \bibnamefont {White}}, \ and\
  \bibinfo {author} {\bibfnamefont {M.~L.}\ \bibnamefont {Brongersma}},\
  }\bibfield  {title} {\enquote {\bibinfo {title} {Nonresonant enhancement of
  spontaneous emission in metal-dielectric-metal plasmon waveguide
  structures},}\ }\href@noop {} {\bibfield  {journal} {\bibinfo  {journal}
  {Phys. Rev. B}\ }\textbf {\bibinfo {volume} {78}},\ \bibinfo {pages} {153111}
  (\bibinfo {year} {2008})}\BibitemShut {NoStop}%
\bibitem [{\citenamefont {Chang}\ \emph {et~al.}(2007)\citenamefont {Chang},
  \citenamefont {S{\o}rensen}, \citenamefont {Hemmer},\ and\ \citenamefont
  {Lukin}}]{chang2007strong}%
  \BibitemOpen
  \bibfield  {author} {\bibinfo {author} {\bibfnamefont {Darrick~E}\
  \bibnamefont {Chang}}, \bibinfo {author} {\bibfnamefont {Anders~S{\o}ndberg}\
  \bibnamefont {S{\o}rensen}}, \bibinfo {author} {\bibfnamefont
  {PR}~\bibnamefont {Hemmer}}, \ and\ \bibinfo {author} {\bibfnamefont
  {MD}~\bibnamefont {Lukin}},\ }\bibfield  {title} {\enquote {\bibinfo {title}
  {Strong coupling of single emitters to surface plasmons},}\ }\href@noop {}
  {\bibfield  {journal} {\bibinfo  {journal} {Phys. Rev. B}\ }\textbf {\bibinfo
  {volume} {76}},\ \bibinfo {pages} {035420} (\bibinfo {year}
  {2007})}\BibitemShut {NoStop}%
\bibitem [{\citenamefont {Akimov}\ \emph {et~al.}(2007)\citenamefont {Akimov},
  \citenamefont {Mukherjee}, \citenamefont {Yu}, \citenamefont {Chang},
  \citenamefont {Zibrov}, \citenamefont {Hemmer}, \citenamefont {Park},\ and\
  \citenamefont {Lukin}}]{akimov2007generation}%
  \BibitemOpen
  \bibfield  {author} {\bibinfo {author} {\bibfnamefont {AV}~\bibnamefont
  {Akimov}}, \bibinfo {author} {\bibfnamefont {A}~\bibnamefont {Mukherjee}},
  \bibinfo {author} {\bibfnamefont {CL}~\bibnamefont {Yu}}, \bibinfo {author}
  {\bibfnamefont {DE}~\bibnamefont {Chang}}, \bibinfo {author} {\bibfnamefont
  {AS}~\bibnamefont {Zibrov}}, \bibinfo {author} {\bibfnamefont
  {PR}~\bibnamefont {Hemmer}}, \bibinfo {author} {\bibfnamefont
  {H}~\bibnamefont {Park}}, \ and\ \bibinfo {author} {\bibfnamefont
  {MD}~\bibnamefont {Lukin}},\ }\bibfield  {title} {\enquote {\bibinfo {title}
  {Generation of single optical plasmons in metallic nanowires coupled to
  quantum dots},}\ }\href@noop {} {\bibfield  {journal} {\bibinfo  {journal}
  {Nature}\ }\textbf {\bibinfo {volume} {450}},\ \bibinfo {pages} {402}
  (\bibinfo {year} {2007})}\BibitemShut {NoStop}%
\bibitem [{\citenamefont {Chen}\ \emph {et~al.}(2015)\citenamefont {Chen},
  \citenamefont {Zenin}, \citenamefont {Leosson}, \citenamefont {Shi},
  \citenamefont {Nielsen},\ and\ \citenamefont
  {Bozhevolnyi}}]{chen2015efficient}%
  \BibitemOpen
  \bibfield  {author} {\bibinfo {author} {\bibfnamefont {Yiting}\ \bibnamefont
  {Chen}}, \bibinfo {author} {\bibfnamefont {Vladimir~A}\ \bibnamefont
  {Zenin}}, \bibinfo {author} {\bibfnamefont {Kristjan}\ \bibnamefont
  {Leosson}}, \bibinfo {author} {\bibfnamefont {Xueliang}\ \bibnamefont {Shi}},
  \bibinfo {author} {\bibfnamefont {Michael~G}\ \bibnamefont {Nielsen}}, \ and\
  \bibinfo {author} {\bibfnamefont {Sergey~I}\ \bibnamefont {Bozhevolnyi}},\
  }\bibfield  {title} {\enquote {\bibinfo {title} {Efficient interfacing
  photonic and long-range dielectric-loaded plasmonic waveguides},}\
  }\href@noop {} {\bibfield  {journal} {\bibinfo  {journal} {Opt. Express}\
  }\textbf {\bibinfo {volume} {23}},\ \bibinfo {pages} {9100--9108} (\bibinfo
  {year} {2015})}\BibitemShut {NoStop}%
\bibitem [{\citenamefont {Weeber}\ \emph {et~al.}(1999)\citenamefont {Weeber},
  \citenamefont {Dereux}, \citenamefont {Girard}, \citenamefont {Krenn},\ and\
  \citenamefont {Goudonnet}}]{weeber1999plasmon}%
  \BibitemOpen
  \bibfield  {author} {\bibinfo {author} {\bibfnamefont {Jean-Claude}\
  \bibnamefont {Weeber}}, \bibinfo {author} {\bibfnamefont {Alain}\
  \bibnamefont {Dereux}}, \bibinfo {author} {\bibfnamefont {Christian}\
  \bibnamefont {Girard}}, \bibinfo {author} {\bibfnamefont {Joachim~R}\
  \bibnamefont {Krenn}}, \ and\ \bibinfo {author} {\bibfnamefont {Jean-Pierre}\
  \bibnamefont {Goudonnet}},\ }\bibfield  {title} {\enquote {\bibinfo {title}
  {Plasmon polaritons of metallic nanowires for controlling submicron
  propagation of light},}\ }\href@noop {} {\bibfield  {journal} {\bibinfo
  {journal} {Phys. Rev. B}\ }\textbf {\bibinfo {volume} {60}},\ \bibinfo
  {pages} {9061} (\bibinfo {year} {1999})}\BibitemShut {NoStop}%
\bibitem [{\citenamefont {Krenn}\ \emph {et~al.}(2002)\citenamefont {Krenn},
  \citenamefont {Lamprecht}, \citenamefont {Ditlbacher}, \citenamefont
  {Schider}, \citenamefont {Salerno}, \citenamefont {Leitner},\ and\
  \citenamefont {Aussenegg}}]{krenn2002non}%
  \BibitemOpen
  \bibfield  {author} {\bibinfo {author} {\bibfnamefont {Joachim~R}\
  \bibnamefont {Krenn}}, \bibinfo {author} {\bibfnamefont {B}~\bibnamefont
  {Lamprecht}}, \bibinfo {author} {\bibfnamefont {Harald}\ \bibnamefont
  {Ditlbacher}}, \bibinfo {author} {\bibfnamefont {Gerburg}\ \bibnamefont
  {Schider}}, \bibinfo {author} {\bibfnamefont {Marco}\ \bibnamefont
  {Salerno}}, \bibinfo {author} {\bibfnamefont {Alfred}\ \bibnamefont
  {Leitner}}, \ and\ \bibinfo {author} {\bibfnamefont {Franz~R}\ \bibnamefont
  {Aussenegg}},\ }\bibfield  {title} {\enquote {\bibinfo {title}
  {Non--diffraction-limited light transport by gold nanowires},}\ }\href@noop
  {} {\bibfield  {journal} {\bibinfo  {journal} {EPL (Europhysics Letters)}\
  }\textbf {\bibinfo {volume} {60}},\ \bibinfo {pages} {663} (\bibinfo {year}
  {2002})}\BibitemShut {NoStop}%
\bibitem [{\citenamefont {Berini}(2000)}]{berini2000plasmon}%
  \BibitemOpen
  \bibfield  {author} {\bibinfo {author} {\bibfnamefont {Pierre}\ \bibnamefont
  {Berini}},\ }\bibfield  {title} {\enquote {\bibinfo {title}
  {Plasmon-polariton waves guided by thin lossy metal films of finite width:
  Bound modes of symmetric structures},}\ }\href@noop {} {\bibfield  {journal}
  {\bibinfo  {journal} {Phys. Rev. B}\ }\textbf {\bibinfo {volume} {61}},\
  \bibinfo {pages} {10484} (\bibinfo {year} {2000})}\BibitemShut {NoStop}%
\bibitem [{\citenamefont {Zenin}\ \emph {et~al.}(2016)\citenamefont {Zenin},
  \citenamefont {Malureanu}, \citenamefont {Radko}, \citenamefont
  {Lavrinenko},\ and\ \citenamefont {Bozhevolnyi}}]{zenin2016near}%
  \BibitemOpen
  \bibfield  {author} {\bibinfo {author} {\bibfnamefont {Vladimir~A}\
  \bibnamefont {Zenin}}, \bibinfo {author} {\bibfnamefont {Radu}\ \bibnamefont
  {Malureanu}}, \bibinfo {author} {\bibfnamefont {Ilya~P}\ \bibnamefont
  {Radko}}, \bibinfo {author} {\bibfnamefont {Andrei~V}\ \bibnamefont
  {Lavrinenko}}, \ and\ \bibinfo {author} {\bibfnamefont {Sergey~I}\
  \bibnamefont {Bozhevolnyi}},\ }\bibfield  {title} {\enquote {\bibinfo {title}
  {Near-field characterization of bound plasmonic modes in metal strip
  waveguides},}\ }\href@noop {} {\bibfield  {journal} {\bibinfo  {journal}
  {Opt. Express}\ }\textbf {\bibinfo {volume} {24}},\ \bibinfo {pages}
  {4582--4590} (\bibinfo {year} {2016})}\BibitemShut {NoStop}%
\bibitem [{\citenamefont {Brongersma}\ \emph {et~al.}(2000)\citenamefont
  {Brongersma}, \citenamefont {Hartman},\ and\ \citenamefont
  {Atwater}}]{brongersma2000electromagnetic}%
  \BibitemOpen
  \bibfield  {author} {\bibinfo {author} {\bibfnamefont {Mark~L}\ \bibnamefont
  {Brongersma}}, \bibinfo {author} {\bibfnamefont {John~W}\ \bibnamefont
  {Hartman}}, \ and\ \bibinfo {author} {\bibfnamefont {Harry~A}\ \bibnamefont
  {Atwater}},\ }\bibfield  {title} {\enquote {\bibinfo {title} {Electromagnetic
  energy transfer and switching in nanoparticle chain arrays below the
  diffraction limit},}\ }\href@noop {} {\bibfield  {journal} {\bibinfo
  {journal} {Phys. Rev. B}\ }\textbf {\bibinfo {volume} {62}},\ \bibinfo
  {pages} {R16356} (\bibinfo {year} {2000})}\BibitemShut {NoStop}%
\bibitem [{\citenamefont {Maier}\ \emph {et~al.}(2003)\citenamefont {Maier},
  \citenamefont {Kik}, \citenamefont {Atwater}, \citenamefont {Meltzer},
  \citenamefont {Harel}, \citenamefont {Koel},\ and\ \citenamefont
  {Requicha}}]{maier2003local}%
  \BibitemOpen
  \bibfield  {author} {\bibinfo {author} {\bibfnamefont {Stefan~A}\
  \bibnamefont {Maier}}, \bibinfo {author} {\bibfnamefont {Pieter~G}\
  \bibnamefont {Kik}}, \bibinfo {author} {\bibfnamefont {Harry~A}\ \bibnamefont
  {Atwater}}, \bibinfo {author} {\bibfnamefont {Sheffer}\ \bibnamefont
  {Meltzer}}, \bibinfo {author} {\bibfnamefont {Elad}\ \bibnamefont {Harel}},
  \bibinfo {author} {\bibfnamefont {Bruce~E}\ \bibnamefont {Koel}}, \ and\
  \bibinfo {author} {\bibfnamefont {Ari~AG}\ \bibnamefont {Requicha}},\
  }\bibfield  {title} {\enquote {\bibinfo {title} {Local detection of
  electromagnetic energy transport below the diffraction limit in metal
  nanoparticle plasmon waveguides},}\ }\href@noop {} {\bibfield  {journal}
  {\bibinfo  {journal} {Nature Materials}\ }\textbf {\bibinfo {volume} {2}},\
  \bibinfo {pages} {229} (\bibinfo {year} {2003})}\BibitemShut {NoStop}%
\bibitem [{\citenamefont {Pile}\ \emph {et~al.}(2005)\citenamefont {Pile},
  \citenamefont {Ogawa}, \citenamefont {Gramotnev}, \citenamefont {Okamoto},
  \citenamefont {Haraguchi}, \citenamefont {Fukui},\ and\ \citenamefont
  {Matsuo}}]{pile_theoretical_2005}%
  \BibitemOpen
  \bibfield  {author} {\bibinfo {author} {\bibfnamefont {D.~F.~P.}\
  \bibnamefont {Pile}}, \bibinfo {author} {\bibfnamefont {T.}~\bibnamefont
  {Ogawa}}, \bibinfo {author} {\bibfnamefont {D.~K.}\ \bibnamefont
  {Gramotnev}}, \bibinfo {author} {\bibfnamefont {T.}~\bibnamefont {Okamoto}},
  \bibinfo {author} {\bibfnamefont {M.}~\bibnamefont {Haraguchi}}, \bibinfo
  {author} {\bibfnamefont {M.}~\bibnamefont {Fukui}}, \ and\ \bibinfo {author}
  {\bibfnamefont {S.}~\bibnamefont {Matsuo}},\ }\bibfield  {title} {\enquote
  {\bibinfo {title} {Theoretical and experimental investigation of strongly
  localized plasmons on triangular metal wedges for subwavelength
  waveguiding},}\ }\href@noop {} {\bibfield  {journal} {\bibinfo  {journal}
  {Appl. Phys. Lett.}\ }\textbf {\bibinfo {volume} {87}},\ \bibinfo {pages}
  {061106} (\bibinfo {year} {2005})}\BibitemShut {NoStop}%
\bibitem [{\citenamefont {Novikov}\ and\ \citenamefont
  {Maradudin}(2002)}]{novikov_channel_2002}%
  \BibitemOpen
  \bibfield  {author} {\bibinfo {author} {\bibfnamefont {I.~V.}\ \bibnamefont
  {Novikov}}\ and\ \bibinfo {author} {\bibfnamefont {A.~A.}\ \bibnamefont
  {Maradudin}},\ }\bibfield  {title} {\enquote {\bibinfo {title} {Channel
  polaritons},}\ }\href@noop {} {\bibfield  {journal} {\bibinfo  {journal}
  {Phys. Rev. B}\ }\textbf {\bibinfo {volume} {66}},\ \bibinfo {pages} {035403}
  (\bibinfo {year} {2002})}\BibitemShut {NoStop}%
\bibitem [{\citenamefont {Pile}\ and\ \citenamefont
  {Gramotnev}(2004)}]{pile_channel_2004}%
  \BibitemOpen
  \bibfield  {author} {\bibinfo {author} {\bibfnamefont {D.~F.~P.}\
  \bibnamefont {Pile}}\ and\ \bibinfo {author} {\bibfnamefont {D.~K.}\
  \bibnamefont {Gramotnev}},\ }\bibfield  {title} {\enquote {\bibinfo {title}
  {Channel plasmon–polariton in a triangular groove on a metal surface},}\
  }\href@noop {} {\bibfield  {journal} {\bibinfo  {journal} {Opt. Lett.}\
  }\textbf {\bibinfo {volume} {29}},\ \bibinfo {pages} {1069--1071} (\bibinfo
  {year} {2004})}\BibitemShut {NoStop}%
\bibitem [{\citenamefont {Veronis}\ and\ \citenamefont
  {Fan}(2005)}]{veronis2005guided}%
  \BibitemOpen
  \bibfield  {author} {\bibinfo {author} {\bibfnamefont {Georgios}\
  \bibnamefont {Veronis}}\ and\ \bibinfo {author} {\bibfnamefont {Shanhui}\
  \bibnamefont {Fan}},\ }\bibfield  {title} {\enquote {\bibinfo {title} {Guided
  subwavelength plasmonic mode supported by a slot in a thin metal film},}\
  }\href@noop {} {\bibfield  {journal} {\bibinfo  {journal} {Opt. Lett.}\
  }\textbf {\bibinfo {volume} {30}},\ \bibinfo {pages} {3359--3361} (\bibinfo
  {year} {2005})}\BibitemShut {NoStop}%
\bibitem [{\citenamefont {Veronis}\ and\ \citenamefont
  {Fan}(2007)}]{veronis2007modes}%
  \BibitemOpen
  \bibfield  {author} {\bibinfo {author} {\bibfnamefont {Georgios}\
  \bibnamefont {Veronis}}\ and\ \bibinfo {author} {\bibfnamefont {Shanhui}\
  \bibnamefont {Fan}},\ }\bibfield  {title} {\enquote {\bibinfo {title} {Modes
  of subwavelength plasmonic slot waveguides},}\ }\href@noop {} {\bibfield
  {journal} {\bibinfo  {journal} {J. Light. Technol.}\ }\textbf {\bibinfo
  {volume} {25}},\ \bibinfo {pages} {2511} (\bibinfo {year}
  {2007})}\BibitemShut {NoStop}%
\bibitem [{\citenamefont {Huang}\ \emph {et~al.}(2014)\citenamefont {Huang},
  \citenamefont {Seo}, \citenamefont {Sarmiento}, \citenamefont {Huo},
  \citenamefont {Harris},\ and\ \citenamefont
  {Brongersma}}]{huang_electrically_2014}%
  \BibitemOpen
  \bibfield  {author} {\bibinfo {author} {\bibfnamefont {Kevin C.~Y.}\
  \bibnamefont {Huang}}, \bibinfo {author} {\bibfnamefont {Min-Kyo}\
  \bibnamefont {Seo}}, \bibinfo {author} {\bibfnamefont {Tomas}\ \bibnamefont
  {Sarmiento}}, \bibinfo {author} {\bibfnamefont {Yijie}\ \bibnamefont {Huo}},
  \bibinfo {author} {\bibfnamefont {James~S.}\ \bibnamefont {Harris}}, \ and\
  \bibinfo {author} {\bibfnamefont {Mark~L.}\ \bibnamefont {Brongersma}},\
  }\bibfield  {title} {\enquote {\bibinfo {title} {Electrically driven
  subwavelength optical nanocircuits},}\ }\href@noop {} {\bibfield  {journal}
  {\bibinfo  {journal} {Nature Photonics}\ }\textbf {\bibinfo {volume} {8}},\
  \bibinfo {pages} {244} (\bibinfo {year} {2014})}\BibitemShut {NoStop}%
\bibitem [{\citenamefont {Bozhevolnyi}\ \emph {et~al.}(2006)\citenamefont
  {Bozhevolnyi}, \citenamefont {Volkov}, \citenamefont {Devaux}, \citenamefont
  {Laluet},\ and\ \citenamefont {Ebbesen}}]{bozhevolnyi2006channel}%
  \BibitemOpen
  \bibfield  {author} {\bibinfo {author} {\bibfnamefont {Sergey~I}\
  \bibnamefont {Bozhevolnyi}}, \bibinfo {author} {\bibfnamefont {Valentyn~S}\
  \bibnamefont {Volkov}}, \bibinfo {author} {\bibfnamefont {Eloise}\
  \bibnamefont {Devaux}}, \bibinfo {author} {\bibfnamefont {Jean-Yves}\
  \bibnamefont {Laluet}}, \ and\ \bibinfo {author} {\bibfnamefont {Thomas~W}\
  \bibnamefont {Ebbesen}},\ }\bibfield  {title} {\enquote {\bibinfo {title}
  {Channel plasmon subwavelength waveguide components including interferometers
  and ring resonators},}\ }\href@noop {} {\bibfield  {journal} {\bibinfo
  {journal} {Nature}\ }\textbf {\bibinfo {volume} {440}},\ \bibinfo {pages}
  {508} (\bibinfo {year} {2006})}\BibitemShut {NoStop}%
\bibitem [{\citenamefont {Spasenovi{\'c}}\ \emph {et~al.}(2009)\citenamefont
  {Spasenovi{\'c}}, \citenamefont {van Oosten}, \citenamefont {Verhagen},\ and\
  \citenamefont {Kuipers}}]{spasenovic2009measurements}%
  \BibitemOpen
  \bibfield  {author} {\bibinfo {author} {\bibfnamefont {M}~\bibnamefont
  {Spasenovi{\'c}}}, \bibinfo {author} {\bibfnamefont {D}~\bibnamefont {van
  Oosten}}, \bibinfo {author} {\bibfnamefont {E}~\bibnamefont {Verhagen}}, \
  and\ \bibinfo {author} {\bibfnamefont {L}~\bibnamefont {Kuipers}},\
  }\bibfield  {title} {\enquote {\bibinfo {title} {Measurements of modal
  symmetry in subwavelength plasmonic slot waveguides},}\ }\href@noop {}
  {\bibfield  {journal} {\bibinfo  {journal} {Appl. Phys. Lett.}\ }\textbf
  {\bibinfo {volume} {95}},\ \bibinfo {pages} {203109} (\bibinfo {year}
  {2009})}\BibitemShut {NoStop}%
\bibitem [{\citenamefont {Andryieuski}\ \emph {et~al.}(2014)\citenamefont
  {Andryieuski}, \citenamefont {Zenin}, \citenamefont {Malureanu},
  \citenamefont {Volkov}, \citenamefont {Bozhevolnyi},\ and\ \citenamefont
  {Lavrinenko}}]{andryieuski_direct_2014}%
  \BibitemOpen
  \bibfield  {author} {\bibinfo {author} {\bibfnamefont {Andrei}\ \bibnamefont
  {Andryieuski}}, \bibinfo {author} {\bibfnamefont {Vladimir~A.}\ \bibnamefont
  {Zenin}}, \bibinfo {author} {\bibfnamefont {Radu}\ \bibnamefont {Malureanu}},
  \bibinfo {author} {\bibfnamefont {Valentyn~S.}\ \bibnamefont {Volkov}},
  \bibinfo {author} {\bibfnamefont {Sergey~I.}\ \bibnamefont {Bozhevolnyi}}, \
  and\ \bibinfo {author} {\bibfnamefont {Andrei~V.}\ \bibnamefont
  {Lavrinenko}},\ }\bibfield  {title} {\enquote {\bibinfo {title} {Direct
  {Characterization} of {Plasmonic} {Slot} {Waveguides} and {Nanocouplers}},}\
  }\href@noop {} {\bibfield  {journal} {\bibinfo  {journal} {Nano Lett.}\
  }\textbf {\bibinfo {volume} {14}},\ \bibinfo {pages} {3925--3929} (\bibinfo
  {year} {2014})}\BibitemShut {NoStop}%
\bibitem [{\citenamefont {Zenin}\ \emph {et~al.}(2015)\citenamefont {Zenin},
  \citenamefont {Andryieuski}, \citenamefont {Malureanu}, \citenamefont
  {Radko}, \citenamefont {Volkov}, \citenamefont {Gramotnev}, \citenamefont
  {Lavrinenko},\ and\ \citenamefont {Bozhevolnyi}}]{zenin2015boosting}%
  \BibitemOpen
  \bibfield  {author} {\bibinfo {author} {\bibfnamefont {Vladimir~A}\
  \bibnamefont {Zenin}}, \bibinfo {author} {\bibfnamefont {Andrei}\
  \bibnamefont {Andryieuski}}, \bibinfo {author} {\bibfnamefont {Radu}\
  \bibnamefont {Malureanu}}, \bibinfo {author} {\bibfnamefont {Ilya~P}\
  \bibnamefont {Radko}}, \bibinfo {author} {\bibfnamefont {Valentyn~S}\
  \bibnamefont {Volkov}}, \bibinfo {author} {\bibfnamefont {Dmitri~K}\
  \bibnamefont {Gramotnev}}, \bibinfo {author} {\bibfnamefont {Andrei~V}\
  \bibnamefont {Lavrinenko}}, \ and\ \bibinfo {author} {\bibfnamefont
  {Sergey~I}\ \bibnamefont {Bozhevolnyi}},\ }\bibfield  {title} {\enquote
  {\bibinfo {title} {Boosting local field enhancement by on-chip nanofocusing
  and impedance-matched plasmonic antennas},}\ }\href@noop {} {\bibfield
  {journal} {\bibinfo  {journal} {Nano Lett.}\ }\textbf {\bibinfo {volume}
  {15}},\ \bibinfo {pages} {8148--8154} (\bibinfo {year} {2015})}\BibitemShut
  {NoStop}%
\bibitem [{\citenamefont {Krauss}\ \emph {et~al.}(2018)\citenamefont {Krauss},
  \citenamefont {Kullock}, \citenamefont {Wu}, \citenamefont {Geisler},
  \citenamefont {Lundt}, \citenamefont {Kamp},\ and\ \citenamefont
  {Hecht}}]{krauss2018controlled}%
  \BibitemOpen
  \bibfield  {author} {\bibinfo {author} {\bibfnamefont {Enno}\ \bibnamefont
  {Krauss}}, \bibinfo {author} {\bibfnamefont {Rene}\ \bibnamefont {Kullock}},
  \bibinfo {author} {\bibfnamefont {Xiaofei}\ \bibnamefont {Wu}}, \bibinfo
  {author} {\bibfnamefont {Peter}\ \bibnamefont {Geisler}}, \bibinfo {author}
  {\bibfnamefont {Nils}\ \bibnamefont {Lundt}}, \bibinfo {author}
  {\bibfnamefont {Martin}\ \bibnamefont {Kamp}}, \ and\ \bibinfo {author}
  {\bibfnamefont {Bert}\ \bibnamefont {Hecht}},\ }\bibfield  {title} {\enquote
  {\bibinfo {title} {Controlled growth of high-aspect-ratio single-crystalline
  gold platelets},}\ }\href@noop {} {\bibfield  {journal} {\bibinfo  {journal}
  {Crystal Growth \& Design}\ }\textbf {\bibinfo {volume} {18}},\ \bibinfo
  {pages} {1297--1302} (\bibinfo {year} {2018})}\BibitemShut {NoStop}%
\bibitem [{\citenamefont {Ocelic}\ \emph {et~al.}(2006)\citenamefont {Ocelic},
  \citenamefont {Huber},\ and\ \citenamefont
  {Hillenbrand}}]{ocelic2006pseudoheterodyne}%
  \BibitemOpen
  \bibfield  {author} {\bibinfo {author} {\bibfnamefont {Nenad}\ \bibnamefont
  {Ocelic}}, \bibinfo {author} {\bibfnamefont {Andreas}\ \bibnamefont {Huber}},
  \ and\ \bibinfo {author} {\bibfnamefont {Rainer}\ \bibnamefont
  {Hillenbrand}},\ }\bibfield  {title} {\enquote {\bibinfo {title}
  {Pseudoheterodyne detection for background-free near-field spectroscopy},}\
  }\href@noop {} {\bibfield  {journal} {\bibinfo  {journal} {Appl. Phys.
  Lett.}\ }\textbf {\bibinfo {volume} {89}},\ \bibinfo {pages} {101124}
  (\bibinfo {year} {2006})}\BibitemShut {NoStop}%
\bibitem [{\citenamefont {Johnson}\ and\ \citenamefont
  {Christy}(1972)}]{johnson1972optical}%
  \BibitemOpen
  \bibfield  {author} {\bibinfo {author} {\bibfnamefont {Peter~B}\ \bibnamefont
  {Johnson}}\ and\ \bibinfo {author} {\bibfnamefont {R-W\_}\ \bibnamefont
  {Christy}},\ }\bibfield  {title} {\enquote {\bibinfo {title} {Optical
  constants of the noble metals},}\ }\href@noop {} {\bibfield  {journal}
  {\bibinfo  {journal} {Phys. Rev. B}\ }\textbf {\bibinfo {volume} {6}},\
  \bibinfo {pages} {4370} (\bibinfo {year} {1972})}\BibitemShut {NoStop}%
\end{thebibliography}%

\end{document}